\documentclass{aa}
\usepackage{epsfig}

\begin{document}

   \thesaurus{06     
              (08.14.1; 08.11.1; 08.18.1; 08.02.3; 08.16.6; 08.19.4)}  
   \title{The observational evidence pertinent to possible kick mechanisms 
in neutron stars}


   \author{A.A. Deshpande$^1$, R. Ramachandran$^{2,3}$, V. Radhakrishnan$^{1,3}$}

   \offprints{desh@rri.ernet.in}

   \institute{$^1$Raman Research Institute,
              C. V. Raman Avenue, Bangalore - 560 080, India\\
	      $^2$Netherlands Foundation for Research in Astronomy, Postbus 2,
              7990 AA Dwingeloo, The Netherlands. \\
              $^3$Sterrenkundig Instituut, Universiteit van Amsterdam, NL-1098 SJ
              Amsterdam, The Netherlands. \\
              email:desh@rri.ernet.in; ramach@astro.uva.nl; rad@rri.ernet.in}

   \date{Received 4 June 1999 / Accepted ?}

   \authorrunning{Deshpande et al.}

   \titlerunning{Observational evidence for neutron star kick mechanisms}
   \maketitle

\begin{abstract} 
We examine available observations on pulsars for evidence pertaining to
mechanisms proposed to explain the origin of their velocities. We find that
mechanisms predicting a correlation between the rotation axis and the pulsar
velocity are ruled out. Also, that there is no significant correlation between
pulsar magnetic field strengths and velocities. With respect to recent
suggestions postulating asymmetric impulses at birth being solely responsible for
both the spins and velocities of pulsars, single impulses of any duration and
multiple extended duration impulses appear ruled out.
\end{abstract} 

\section{Introduction}
  It is now widely accepted that the velocities observed for pulsars include a
significant component from kicks experienced by the neutron stars in the process
of their formation. The basis for this view is almost totally empirical, with a
variety of different kinds of observations all pointing to the existence of an
impulsive transfer of momentum to the protoneutron star at birth (Shklovskii
1970; Gunn \& Ostriker 1970; van den Heuvel \& van Paradijs 1997).  This implies
an asymmetry in the ejection process, but as yet there is no consensus on any
plausible mechanism for providing such an asymmetry.  Mechanisms suggested range
from hydrodynamical instabilities to those in which asymmetric neutrino emission
is postulated (Burrows 1987; Keil et al.  1996; Horowitz \& Li 1997,
hep-ph/9701214; Lai \& Qian 1998, astro-ph/9802344; Spruit \& Phinney 1998).

As far as the latter class is concerned, it appears, and very reasonably
so, that if the neutrinos can impart momentum to the matter, the reverse
must also happen, and the thermal equilibrium of the matter must
necessarily destroy any incipient asymmetry in the neutrinosphere. And
any asymmetry developed above the neutrino-matter decoupling layer, is by
definition incapable of imparting any momentum to the matter (Bludman 1998,
{\it private communication}).

Whatever the operative mechanism for creating the asymmetry, it is an
important and pertinent question to ask if the resulting direction is a
random one, or connected with some basic property of the protoneutron
star. Two such essential vectors associated with the core of the
collapsing star are its rotational and magnetic axes, and both have been
invoked in mechanisms proposed in the literature (Harrison \& Tademaru
1975a,b; Burrows \& Hayes 1996; Kusenko \& Segre 1996). Any
such mechanism that is postulated to provide the asymmetry must
leave its signature in the direction and magnitude of the imparted
velocity, thus enabling a possible test of the theory by comparison with
observations.

An important recent investigation in this connection is that of Spruit and
Phinney (1998). They argue strongly that the cores of the progenitors of
neutron stars cannot have the angular momentum to explain the rotation of
pulsars and propose birth kicks as the origin of their spins.  These
authors do not specify any particular physical process as responsible for
the ``kick", but emphasize that unless its force is exerted exactly head-on
it must also cause the neutron star to rotate. As both the velocity and
the spin of the neutron star have a common cause according to this
hypothesis, it is not unreasonable to expect testable correlations as we
shall discuss a little later. Independently, Cowsik (1998) has also 
advanced a similar common origin for the proper motion and spin of 
pulsars. The first suggestion of this possibility was by Burrows et al. (1995)

The quantities on which comparisons with observations can be made are the
direction and magnitude of the proper motion, the projected direction of
the magnetic axis, the magnitude of the magnetic field, the direction of
the rotation axis and the initial period of rotation. Of these it is only
the last that is not accessible to observation. We are left with five
quantities which may be interrelated, depending on the mechanism which
causes the asymmetry.  Several types of correlations between these
quantities have been sought, and even claimed in the past, motivated by
suggestions of possible kick mechanisms. Our approach in this
investigation is to examine without prejudice an enlarged body of pulsar
data now available for any correlations which could support or rule out
various suggested explanations.

As is widely practised, we shall assume that $\alpha$, the angle the magnetic
axis makes with respect to the rotational axis, and $\beta$, to the line of
sight, can both be derived from accurate measurements of the core-component
widths and the sweep of the linear polarisation through the pulse window. The
intrinsic angle of polarisation at the point of inflexion of the sweep then
gives us the projection of the rotational axis, and $\alpha + \beta = \zeta$ the
complement of the angle it makes to the plane of the sky. We shall also assume
that the observed proper motion is due only to the kick received at birth, and
return later to a discussion of contributions to the velocity from motion of the
progenitor in a binary system, or from its runaway velocity from a previous
disruption.

The simplest models from the point of view of testability are those which
predict an acceleration strictly along the rotational axis such as the
rocket mechanism of Harrison and Tademaru (1975a; 1975b). The simplicity is due to
projection on the sky plane not affecting the alignment expected of the
rotation axis and the velocity direction (Morris et al. 1976). They argued
that the direction of the spin axis projected on the plane of the sky must
be the same as that of the rotation axis. Based on thirteen available
samples at that time, they concluded that the acceleration mechanism
suggested by Harrison \& Tademaru is not supported by observations. A
similar conclusion was reached by Anderson \& Lyne (1983).

 In this paper, we use high quality polarisation and proper motion observations
on a larger sample to test for preferential alignment of the pulsar velocity
with either the rotation or the magnetic axis.

\section{Sample selection}
We have carefully selected for our study a sample of 29 pulsars for which we
estimated the intrinsic position angle (IPA) from observations available in the
literature (Morris et al. 1981; McCulloch et al. 1978; Manchester et al. 1980;
Xilouris et al. 1991). The transverse velocities were computed from the proper
motion measurements of Lyne, Anderson \& Salter (1982) and Harrison, Lyne \&
Anderson (1993), along with distances estimated using the electron density
distribution model of Taylor \& Cordes (1993). It may be noted that the distance
information is required not only for calculating velocities, but also for
calculating proper motion direction, as correction for differential galactic
rotation needs to be incorporated. The selected list of pulsars together with
the calculated directions of the intrinsic position angles and the proper
motions are given in table \ref{table:tab1}.

\begin{table}
\begin{center}
\begin{tabular}{l|r|l|r|l} \hline
{\bf Name}  &  {\bf IPA} & {\bf $\Delta$IPA}  & {\bf PM$_{\rm dir}$} & {\bf
$\Delta$PM$_{\rm dir}$} \\ 
           & (~$^{\circ}$) & (~$^{\circ}$) &(~$^{\circ}$) &(~$^{\circ}$) \\ \hline
   B0136+57   &     30 &  10 &   210 &  19     \\
   B0301+19   &     22 &   4 &   171 &  13     \\
   B0329+54$^{\dagger}$  &     10 &   4 &   127 &   4     \\
   B0450-18$^{\dagger}$  &     38 &   5 &    28 &  27     \\
   B0450+55   &     56 &  10 &   108 &   7     \\
   B0523+11   &     80 &  13 &    98 &  16     \\
   B0540+23   &    125 &  15 &    58 &  25 \\
   B0736-40$^{\dagger}$  &    156 &   5 &   309 &  10 \\
   B0740-28   &     90 &  15 &   270 &   2     \\
   B0809+74$^{\dagger}$  &    167 &   7 &   163 &  11     \\
   B0818-13   &     50 &   5 &   164 &  19     \\
   B0823+26   &     59 &   3 &   146 &   2 \\
   B0833-45   &     35 &  10 &   306 &   3     \\
   B0834+06$^{\dagger}$  &     76 &  15 &     2 &   6     \\
   B1133+16$^{\dagger}$  &    100 &   5 &   344 &   1     \\
   B1237+25   &    160 &   8 &   292 &   3     \\
   B1449-64$^{\dagger}$  &    140 &   8 &   217 &   4     \\
   B1508+55   &    161 &   5 &   227 &   3 \\
   B1706-16$^{\dagger}$  &    123 &   7 &    27 &  18 \\
   B1818-04$^{\dagger}$  &     67 &  13 &     6 &   7 \\
   B1905+39   &     68 &  15 &    45 &  12     \\
   B1929+10$^{\dagger}$  &     53 &  10 &    68 &   4 \\
   B1933+16$^{\dagger}$  &    150 &  15 &   175 &  16 \\
   B2020+28$^{\dagger}$  &    161 &   5 &   215 &  17     \\
   B2021+51   &     23 &   5 &    19 &  14     \\
   B2045-16$^{\dagger}$  &      0 &   3 &   117 &  14 \\
   B2154+40   &    104 &   6 &    99 &   4     \\
   B2217+47   &     58 &  10 &   202 &  19     \\
   B2351+61$^{\dagger}$  &    124 &  20 &    75 &   8 \\
\end{tabular}
\end{center}
\caption[]{List of selected pulsars with accurately determined directions of
intrinsic position angles (IPA) and proper motions. The second and third columns
give the IPA and its measurement error. The fourth and the fifth columns give
the direction of PM and its measurement error. \\ $^{\dagger}$Cases where the
Intrinsic position angles have an ambiguity due to orthogonal flips.}
\label{table:tab1}
\end{table}

It is worth mentioning here that most of the data used by Anderson \& Lyne
(1983) were from Morris et al. (1979; 1981) where the position angle of
polarisation was measured at the centre of the average pulse profiles.  This, as
we have since found, can be a source of significant error in the estimation of
the IPA. We have included in the present comparison only those objects where the
point of inflexion in the position angle sweep could be clearly identified.

\begin{figure}
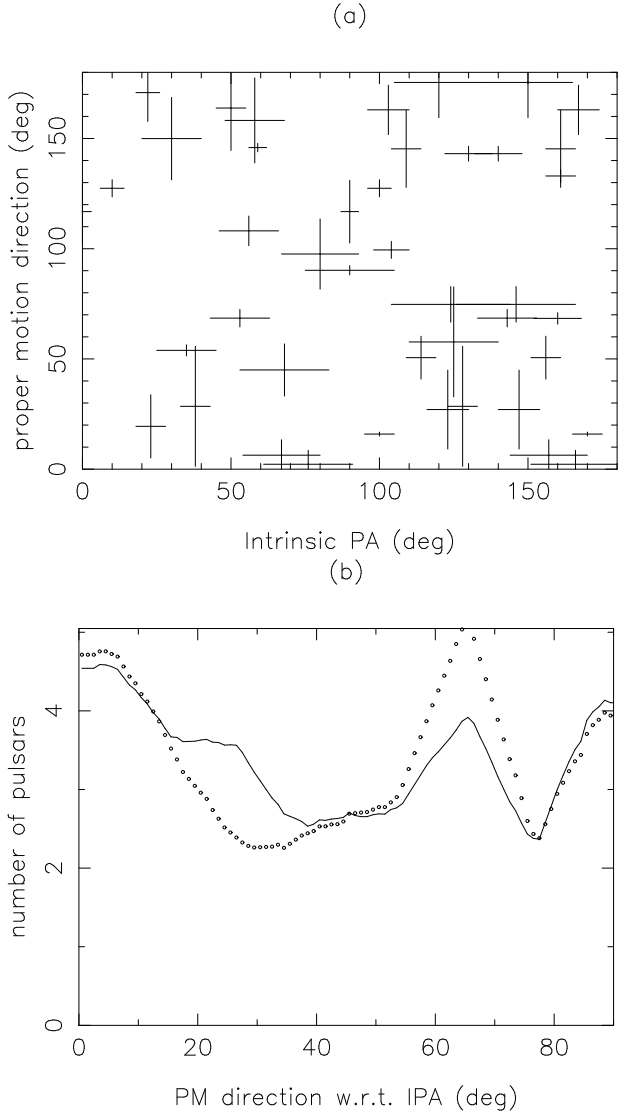

\epsfig{file=H1583_fig1a.ps,width=2.9in,angle=-90}
\epsfig{file=H1583_fig1b.ps,width=2.9in,angle=-90}
\caption[]{{\bf (a)} Direction of proper motion vs intrinsic position angle in
the plane of the sky.
{\bf (b)} Probability distribution of the difference
between these two angles. Points have been weighted proportional to the
reciprocal of the measurement errors, and the distribution has
been smoothed with a window of 10 degree width. The dotted curve shows the
distribution when orthogonal flips in the polarisation position angles are
{\it not} accounted for. The sample used here corresponds to 29 pulsars
selected on the basis that the measurement error in the angle differences
does not exceed 30$^o$.}
\label{fig:fig1}
\end{figure}

\section{Velocity parallel to rotation axis}
If the velocity vector of the neutron star is along its rotation
axis, then it means that either the underlying mechanism itself produces 
acceleration  along the rotation axis, or that the time scale over which the
asymmetry is produced during the supernova explosion is much longer
than the rotation period of the star, thus averaging out the azimuthal
component. In both cases, the direction of the proper
motion vector in the plane of the sky should be the same as the projected 
direction of the rotation axis. 

Figure \ref{fig:fig1}a shows the distribution of the proper motion 
directions and the IPAs for 29 pulsars. The values of the IPAs have been 
obtained after correcting for interstellar Faraday rotation. In the 
lower panel the distribution of the difference between these two angles 
is shown. Pulsars (14 out of the 29) with orthogonal flips in the 
polarisation position angles are included with the IPA assigned two 
possible values 90$^o$ apart. 
Since the observational errors are not the same for all the points, each
point has been weighted proportional to the reciprocal of the error in the
angle difference. The distribution has been smoothed with a 10 degrees wide 
smoothing window. As is evident from this figure, there is no 
significant peak  at either 0 or 90 degrees. 
So, even with the improved data set, any significant matching between the 
rotation axis and the direction of proper motion is not seen.

\section{`Kicks' along the magnetic axis of the star}
We consider next mechanisms that would predict one or more momentum impulses
directed along the magnetic axis and proportional to the strength of the
field. In such a scenario, the resultant direction of the motion would depend on
the duration of the impulse as compared to the (unknown) period of rotation at
the epoch of the impulse. Short impulses would accelerate the neutron star along
the instantaneous (and unknown) direction of the magnetic axis, and long
impulses, with net duration longer than $\sim50\%$ of the rotation period of the
star, would result in motion increasingly along the rotational axis due to
averaging, and of a magnitude now proportional to $\cos\alpha$ times the
magnetic field strength, $B$. We have already shown that there is no correlation
between the directions of the proper motions and the projected rotation
axes. This appears to rule out slow impulses and leave only the case of short
impulses along the magnetic axis to be considered.  But before going on to the
short-impulse case, let us look at a possible correlation between the magnetic
field strength and the magnitude of the observed velocity for long duration
impulses.

\subsection{Field strength vs velocity}
The fact that we see a given pulsar means that the angle between the rotation
axis and the plane of the sky is $[90-(\alpha + \beta)]^o$, where $\beta$ is the
minimum angle between the line of sight and the magnetic axis (impact angle). 
Therefore, if the magnitude of the extended impulse is proportional to the
strength of the magnetic field, the net transverse velocity of the pulsar must be 
proportional to $B\cos(\alpha)\sin(\alpha+\beta)$. 
In most pulsars, $\beta$
is much smaller than $\alpha$ (and often its sign is not known) and 
the observed transverse
velocity should be approximately proportional to $B\sin(2\alpha)/2$, i.e.,
the transverse velocity will at maximum have half of the potential kick 
when $\alpha=45$ degrees. 

Incorporating all these considerations, we examined 
$V_{\rm pm}/\sin(\alpha + \beta)$ vs $B\cos\alpha$ for a 
sample of 44 pulsars for which the relevant
quantities are known reliably (for example, pulsars with more than 50\% uncertainty in 
proper motion have been excluded). 
The values of $\alpha$, $\beta$ used here 
are taken from Rankin(1993).
\footnote{Use of $\alpha$, $\beta$ values from Lyne \&
Manchester (1988) leads to a slight worsening of the correlation.}. 
 Any correlation present does not appear to be
statistically robust. 
For normal pulsars (rotation periods longer than 25 milliseconds),
the correlation coefficient is $0.35\pm 0.15$ (see Fig. 2). 
This seems to improve marginally, to $0.65\pm 0.25$ if pulsars with 
large errors in their proper motion are weighted down, but the correlation is, 
by no means, statistically significant.
This is in agreement with the earlier work of Lorimer, Lyne \& Anderson (1995),
Birkel \& Toldra (1997) and Cordes \& Chernoff (1998).

\subsection{The case of single short-lived kicks}
As pointed out above, the direction of the magnetic axis projected on the plane
of the sky at the time of explosion is an unknown quantity. If the rotation
axis is located in the plane of the sky, then the projected magnetic axis in
the plane of the sky will be within $\pm\alpha$ of the rotation axis.
However, when the direction of the rotation axis is oriented close to the
line of sight, the varying angle between the projected magnetic and
rotation axes will always exceed the above range. 
%


   To assess this issue in detail,  
   let us define the z-axis as pointing towards the observer,
   then x-y is the plane of the sky. Choose the x-axis to be
   along the projected rotation axis on the plane of the sky
   (i.e. aligned with the intrinsic PA).
     Then it is easy to see that, as the star rotates, the
   components of the kick imparted along the instantaneous
   magnetic-field-direction (and proportional to B,
   the magnetic-field strength) are given by

\begin{equation}
V_x = k.B \left[\cos\alpha\;\sin(\alpha+\beta) 
                 - \sin\alpha\;\cos(\alpha+\beta)\;\cos\phi\right]
\end {equation}
\begin{equation}
V_y = k.B [\sin\alpha\;\sin\phi]
\end {equation}
\begin{equation}
V_z = k.B [\cos\alpha\;\cos(\alpha+\beta) 
                 + \sin\alpha\;\sin(\alpha+\beta)\;\cos\phi]  
\end {equation}
     where k is a constant of proportionality,
       and $\phi$ is the rotation phase (like the pulse
           longitude; $\phi$=0 for the closest angle between
           the magnetic-field direction and the sight-line).

   Then it follows that the angle ($\theta$) which the proper-motion 
   direction would make with respect to the intrinsic
   PA direction is given by

\begin{equation}
   \tan\theta =\frac{ \sin\alpha\;\sin\phi}
               {\cos\alpha\;\sin\zeta - \sin\alpha\;\cos\zeta\;\cos\phi}
\end{equation}

\begin{figure}
\epsfig{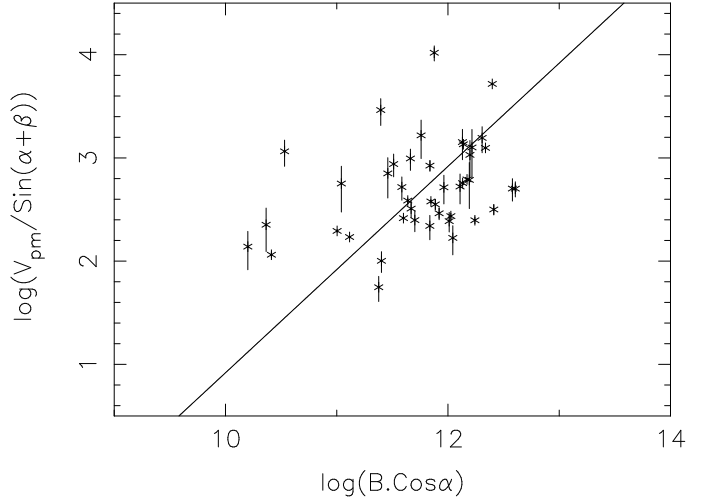}
\caption[]{A plot of $V_{pm}/\sin(\alpha+\beta)$ versus $B\cos(\alpha)$, where
$V_{pm}$ is in km/s and $B$ in Gauss. The straight line corresponds to the linear
best-fit dependence. See text for details.}
\label{fig:fig2}
\end{figure}

       where $\zeta = \alpha+\beta$.  This is exactly the expression describing
   the sweep of the position angle of pulsar polarisation in the model of
   Radhakrishnan \& Cooke (1969).  The magnitude of the transverse velocity is
   then simply $ V_{xy} = \sqrt{V_x^2 + V_y^2}$.  With this understanding, we
   explore the following two approaches.

     {\bf a)} We take the IPA and proper-motion direction at their face value.
         Their relative angle (i.e. the difference) should allow us to compute
         $\phi$ (never mind the sense of rotation)--- giving us 8 possible
         solutions, of which we need to consider only the 4 independent ones
         (say $\phi_1$ to $\phi_4$) given the symmetry in the
         problem\footnote{The allowance for any orthogonal-flips in the
         polarisation position angle makes the possible solutions 8 instead of
         the 4 that come from simple projection considerations, and the
         left-right symmetry.}.  We used these values, $\phi_1$ to $\phi_4$, to
         estimate the expected magnitudes of the proper-motions and compared
         them with the measured values\footnote{Note that the sample size here
         is smaller than that shown in Fig. 2, due to limited measurements of
         IPA \& proper-motion direction.}.  Even in the best case any
         correlation present was not significant and hence no clear inference
         was possible.\\

    {\bf  b)}  In a second more statistical approach, we compute
         for each of the sample pulsars, 
         the distribution of the relative angle $\angle(V_{pm} -IPA)$
         based on the known geometry (along with the above equations)
         and by assuming that $\phi$ is distributed uniformly over its range.
         A combined distribution computed this way for a single short impulse
         is not significantly different from
         the observed distribution in Fig. 1.
         Assuming impulses of longer durations leads to a significantly
         greater expectation of alignment which is not seen. 
         The observations therefore
         suggest that if impulses are along the magnetic axis, they must be
         confined to a small fraction of the rotational phase cycle of the star.

\section{Birth kicks as the origin of pulsar rotation}
        The mechanisms examined so far assumed `radial kicks' which did not
        affect the rotation rate of the star.  As noted already, Spruit \&
        Phinney (1998, hereafter SP) and Cowsik (1998) have explored the
        possibility of non-radial random kicks which would impart both net
        linear and angular momenta to the star.  Noting the significance of this
        mechanism, we have looked at this issue in some detail.

        SP give the results of their simulations for a case of 4 random
        non-radial kicks that impart velocities and spins of magnitudes such as
        observed.  We examine this model over a range of parameters, such as the
        number of momentum impulses, their magnitudes and durations.  For each
        combination of the parameters, we examine the resultant distribution of
        the angle that the apparent motion would make relative to the projection
        of the rotation axis of the star in the plane of the sky. We of course
        compare the resultant distribution with what is observed (Fig. 1).  We
        select for our examination only the subset that most corresponds to the
        expected dispersion in the velocity and in the rotation rates of the
        sample.

\begin{figure}
\epsfig{file=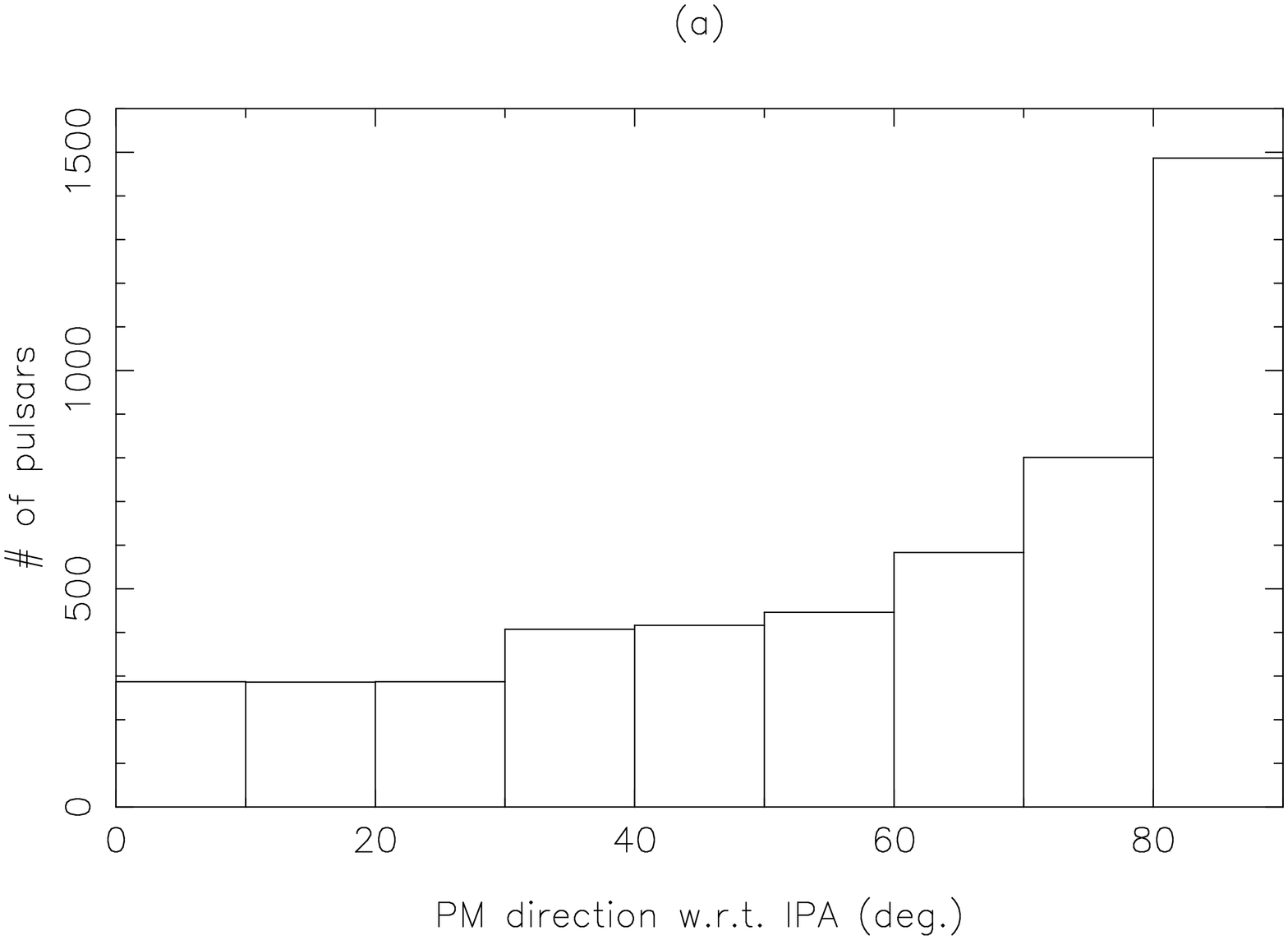,width=2.6in,angle=-90}
\epsfig{file=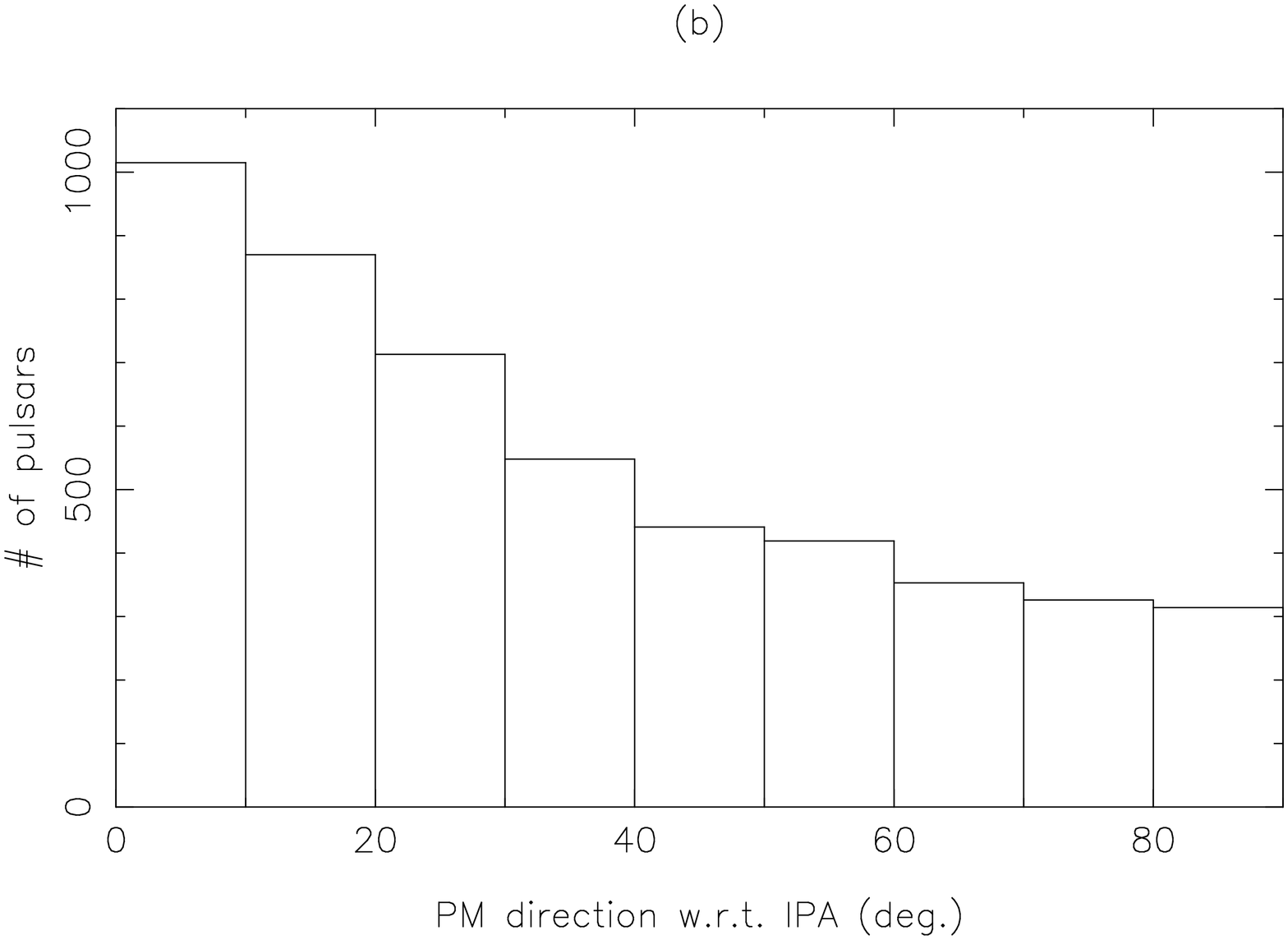,width=2.6in,angle=-90}
\caption[]{The expected distributions of the angles between the proper motion
and the spin axis projected on the sky. {\bf (a)} For a single non-radial impulse of
any duration, there is a significant bias towards 90 degrees. {\bf (b)} In the case
of of a number of long duration impulses, the bias is towards zero degrees.}
\label{fig:simulfig}
\end{figure}

         From our numerical simulation\footnote{5000 sample stars considered in
         each case.}, we find the following.  As noted by SP, a single impulse
         gives linear motion in a plane perpendicular to the rotation axis. In
         the plane of the sky, the projected direction of the star motion
         relative to that of the spin axis would be expected to show a
         significant bias towards 90$^o$, as illustrated in Fig.
         \ref{fig:simulfig}a. This is $not$ observed, as seen in Fig.
         \ref{fig:fig1}. Hence, single impulses being responsible for both the
         birth velocity and spin can be ruled out. This holds irrespective of
         their duration as the rotation axis itself is defined by the action of
         the impulse. The strongest basis for this conclusion comes from the
         knowledge of the particular orientation angles of the pulsars in our
         sample.

         Hence, in such models, there must be two or more randomly directed
         impulses to be consistent with the observations.  As might be expected,
         the strength of the impulses required to obtain the observed range of
         velocities and spin rates will be weaker (by a factor $\sqrt{N}$) for
         an increasing number of impulses ($N$).  The velocity dispersion as a
         function of the duration of a momentum impulse of a given magnitude
         remains constant up to a certain critical duration ($\tau_c$), above
         which the azimuthal averaging reduces the resultant velocities.
         Following SP, if we assume the impact radius (radius of the
         protoneutron star) to be three times the radius of the neutron star, we
         find $\tau_c$ to be about 9 times the corresponding resultant rotation
         period of the star, as also pointed out by SP.  The velocity dispersion
         levels off (at a reduced value) when the impulse duration is much
         larger than $\tau_c$. In contrast, the resultant spin rate is
         independent of the impulse duration but varies linearly with its
         average magnitude ($I$). Thus, the product $(I\times\tau_c)$ is a
         constant that depends on the number of impulses and the star's mass and
         its moment of inertia.

         For impulse durations ($\tau$) much smaller than $\tau_c$, both the
         linear and angular momenta grow as $\sqrt{N}$ but the angle between
         them becomes random.  However, for relatively longer-duration impulses
         a significant preference of the direction of the linear momentum
         develops towards the spin axis, which itself is evolving. This bias is
         illustrated in Fig. \ref{fig:simulfig}b. Since the data on the
         relative inclinations do not show such a preference, we conclude that
         the impulse durations must be shorter or equal to the corresponding
         critical duration $\tau_c$. This also means that no significant
         reduction\footnote{ the reduction in the net velocity/spin due to the
         randomness in the impulse direction will continue to exist.}  in the
         expected net velocities occurs due to azimuthal averaging during
         rotation and therefore somewhat weaker impulses can account for the
         observed velocities. However, the corresponding spin rate would then be
         smaller than with long-duration impulses.  If the duration of each kick
         in the 4-kick situation considered by SP is 0.32 seconds, we find that
         about 55\% of the population should appear to have relative
         inclinations between 0-30$^o$ compared to about 20\% in the
         60$^o$-90$^o$ range. Even after due allowance for possible selection
         effects etc., we find that the expected distribution of the relative
         directions is far from what is observed. Hence, both the duration and 
         the magnitude of the impulses in the example considered by SP are to be
         reduced by factors of 5 or more and of 2, respectively.

\section{Discussion}
We have shown in this work that given the present sample of radio pulsars for
which we have reliable proper motion and polarisation measurements, no
significant correlation exists between the magnetic field strength and the
magnitude of the spatial velocities of pulsars, or between the projected
directions of the rotation axis (and/or the magnetic axis) and the direction of
the proper motion vector. This has fundamental implications for the mechanisms
producing the asymmetric supernova kick velocities, as the observations do not
support any mechanism producing net kick velocities parallel to the rotation
axis. This rules out therefore, momentum impulses of any duration along the
rotation axis, and any long duration (compared to the rotation period) impulses
along any one fixed axis, for example, the magnetic axis.

Among the most elegant suggestions to explain the origin of kick speeds and the
initial rotation periods of pulsars are the ones suggested by Spruit \& Phinney
(1998) and Cowsik (1998). It has been possible to quantitatively assess the
expectation regarding the distribution of proper-motion directions using the
framework of the SP model. Our simulations and comparison of the results with
the observations show that single impulses are ruled out, but not 2 or more
impulses of relatively short duration. The durations have to be short enough not
to cause any significant azimuthal averaging of the radial component of the
impulse. The same conclusions should also apply to the model of Cowsik (1998).
However, his expectation of an inverse correlation between velocity and initial
rotation period (also applicable to SP) can not be tested readily.

In the above analysis, we have ignored an important aspect of the evolutionary
history of pulsars. As we know, a good fraction of massive stars in the sky are
in binary or multiple systems. This implies that almost by definition, a
considerable fraction of pulsars are born in binary systems. Most of them get
disrupted during the first supernova explosion in the binary. Therefore, the
spatial velocities of such pulsars must still retain some memory of their binary
origin.  The possible contribution to the post-explosion speeds from the
pre-explosion orbital-motion can be appreciable (Radhakrishnan \& Shukre 1985;
Bailes 1989), particularly noting that a predominant number of pulsars seem to
have speeds of the order of only about 200 km/sec (Hansen \& Phinney 1997;
Blaauw \& Ramachandran 1998). Moreover, the analysis of Deshpande et al. (1995)
shows that a considerable fraction of pulsars may be born at large heights from
the galactic plane, suggesting that a considerable fraction of pulsars are born
in binary systems, which have {\it run away} from the plane. It must be
emphasized therefore that our conclusions above are valid only if the velocities
are derived solely from natal kicks.

To estimate this `contamination' from the progenitor orbital velocities, an
identification of an origin in OB-associations for as many pulsars as possible
might throw much light. In some of these cases, the pulsar progenitor may
perhaps be identified with the progenitor of a runaway OB star. We thank the
referee for pointing out this possibility, as also the inadequacy of the present
proper motion determinations for such accurate backtracking; and for emphasizing
that for this problem, increased accuracy of the proper motions of known pulsars
is more important than increasing the sample of known pulsars. More accurate
information on the statistics of orbital velocities in the pre-explosion stage
of massive double star systems would also be very important.

An interesting case of observational evidence relating to the direction of the
birth-kick is the recent work by Wex et al. (1999, private
communication). Through a detailed modelling of the pre-explosion binary
progenitor of PSR B1913+16, they find that the direction of the kick velocity
must have been almost opposite to that of the orbital velocity of the exploding
component to have not disrupted the system.  In such binary systems, one expects
the rotational angular momentum of the star to be parallel to that of the binary
orbit due to the evolutionary history. The kick velocity could therefore not
have been along the rotation axis of the star.

   An orbital velocity contribution can be viewed as from a `kick', but one
which is radial, i.e. it does not contribute to the spin of the star.  It is
easy to see that such a contribution to the velocity can influence the direction
of the net motion, significantly so when its magnitude is comparable to or
greater than the contribution from the natal kicks. In such cases, the resultant
proper motion direction (with respect to the spin axis) should become more and
more random as the relative contribution from the orbital motion increases. We
have verified this expectation through simulation by including an initial
velocity component in a random direction and of a varying magnitude. In all the
cases discussed earlier, where the relative directions of the proper motion were
biased towards the projection of the spin axis (or orthogonal to it), we see, as
would be expected, a significant reduction in the bias. In fact, when the two
possible contributions to the motion are about equal, the expected distribution
of the relative directions becomes indistinguishable from the observed one
(Fig.\ref{fig:fig1})! This is so, independent of the kick durations, thus not
necessarily requiring short duration kicks. The role of the duration of kicks is
limited to only deciding whether azimuthal averaging occurs or not. Long
duration kicks will only reduce the net contribution of the natal kicks to the
proper motion and will no longer be relevant in deciding the direction of the
`net' motion.

Our conclusions are therefore,
\begin{enumerate}
\item Mechanisms predicting a correlation between the rotation axis and the
      pulsar velocity are ruled out by the observations. This includes single
      long duration radial kicks along any fixed axis of the star.
\item There is no significant correlation between the magnetic field strength
      and the velocity.
\item If asymmetric acceleration at birth is responsible for both the rotation
      and the velocity of the pulsar, {\it the observations rule out single
      impulses of any duration and multiple extended duration impulses.}
\item The above conclusions lose their significance if there is a substantial
      contribution to pulsar velocities from the orbital (or runaway) motion
      of the progenitor.
\end{enumerate}

\begin{acknowledgement}
We are grateful to the referee Adrian Blaauw for his valuable comments.
\end{acknowledgement}

\end{document}